\documentclass[twocolumn,reprint,
 superscriptaddress,
 bibnotes,
 amsmath,amssymb,
 aps,
pre,
floatfix,
]{revtex4-2}

\usepackage{graphicx}
\usepackage{dcolumn}
\usepackage{bm}

\DeclareMathOperator{\erfc}{erfc}

\DeclareMathOperator{\erf}{erf}
\DeclareMathAlphabet{\mathdutchcal}{U}{dutchcal}{m}{n} 
\SetMathAlphabet{\mathdutchcal}{bold}{U}{dutchcal}{b}{n}
\DeclareMathAlphabet{\mathdutchbcal}{U}{dutchcal}{b}{n} 

\usepackage{xr}
\usepackage{lipsum}  
\usepackage{color}

\begin{document}

\title{Local Time Statistics and Permeable Barrier Crossing: from Poisson to Birth-Death Diffusion Equations%
}

\author{Toby Kay}
 \email{t.kay@imperial.ac.uk}
\affiliation{Department of Chemical Engineering, Imperial College London\\
London SW7 2AZ, United Kingdom}
\affiliation{
School of Engineering Mathematics and Technology, University of Bristol\\
Bristol, BS8 1UB, United Kingdom 
} 
\author{Luca Giuggioli}%
 \email{Luca.Giuggioli@bristol.ac.uk}
\affiliation{
School of Engineering Mathematics and Technology, University of Bristol\\
Bristol, BS8 1UB, United Kingdom 
}
\date{\today}

\begin{abstract}
    Barrier crossing is a widespread phenomenon across natural and engineering systems. While an abundant cross-disciplinary literature on the topic has emerged over the years, the stochastic underpinnings of the process are yet to be linked quantitatively to easily measurable observables.  We bridge this gap by developing a microscopic representation of Brownian motion in the presence of permeable barriers that allows to treat barriers with constant asymmetric permeabilities. Our approach relies upon reflected Brownian motion and on the crossing events being Poisson processes subordinated by the local time of the underlying motion at the barrier. Within this paradigm we derive the exact expression for the distribution of the number of crossings, and find an experimentally measurable statistical definition of permeability. We employ Feynman-Kac theory to derive and solve a set of governing birth-death diffusion equations and extend them to the case when barrier permeability is constant and asymmetric. As an application we study a system of infinite, identical and periodically placed asymmetric barriers for which we derive analytically effective transport parameters. This periodic arrangement induces an effective drift at long times whose magnitude depends on the difference in the permeability on either side of the barrier as well as on their absolute values. As the asymmetric permeabilities act akin to localised ``ratchet'' potentials that break spatial symmetry and detailed balance, the proposed arrangement of asymmetric barriers provides an example of a noise-induced drift without the need to time-modulate any external force or create temporal correlations on the motion of a diffusing particle. By placing only one asymmetric barrier in a periodic domain we also show the emergence of a non-equilibrium steady state.
\end{abstract}

\maketitle

\section{Introduction}

Biological and man-made systems are replete with spatial heterogeneities that either facilitates or hinders the random movement of agents or particles. A permeable barrier represents one such example whereby the motion statistics of a particle is reduced, while its lifetime remains unaltered, leading to the so-called inert interactions \cite{sarvaharman2023particle,giuggioli2023multi}.
Examples can be found across scales and disciplines, from electrochemical species diffusing through multi-layer electrodes \cite{diard2005one,freger2005diffusion,ngameni2014derivation}, water transport in rock pores \cite{song2000determining} and drug delivery in the epidermis \cite{siegel1986laplace,pontrelli2007mass,todo2013mathematical} to rough terrains or unequal landscapes  affecting animal dispersal \cite{beyer2016you,assis2019road,kenkre2021theory} and the diffusion of water molecules in heterogeneous media for magnetic imaging techniques \cite{grebenkov2014exploring1,grebenkov2014exploring2}. Permeable barriers also play an important role in regulating the flux of biochemicals between spatial regions in cells \cite{phillips2012physical}, such as the bilayer plasma membrane of eukaryotes \cite{kenkre2008molecular,kusumi2005paradigm,nikonenko2021ion} and the electrical gap junctions in neurons \cite{evans2002gap,connors2004electrical}. 

Quantifying random movement in the presence of permeable barriers is often tackled macroscopically, by imposing on the diffusion equation an interface condition that accounts for the barrier \cite{tanner1978transient,powles1992exact,moutal2019diffusion}, or microscopically via random walks, where the barrier alters the movement between two specific lattice sites \cite{powles1992exact,kosztolowicz1998continuous,kosztolowicz2001random,kenkre2008molecular,aho2016diffusion,kay2022diffusion}. However, neither approach addresses how the  underlying Brownian motion is affected by the permeable barrier and how it leads to a boundary value problem (BVP) for the associated Smoluchowski or diffusion equation. Examples that link stochastic description and BVP abound, e.g. the Skorokhod equation which generates reflected BM \cite{skorokhod1961stochastic,skorokhod1962stochastic,ito1963brownian,ito1996diffusion,freidlin1985functional,saisho1987stochastic} for the perfectly reflecting BVP, elastic (partially reflecting) BM \cite{feller1954diffusion,ito1996diffusion,grebenkov2006focus,grebenkov2020paradigm} which is associated with the radiation/Robin BVP, and sticky BM \cite{feller1952parabolic,ito1963brownian,ito1996diffusion,bressloff2023close} for the slowly reflecting BVP. 

In this context, the snapping out BM has been proposed recently as an appropriate stochastic representation of the process of crossing permeable barriers. Originally conceptualized by Lejay \cite{lejay2016snapping}, snapping out BM is constructed by sewing together different excursions of elastic BM on either side of the barrier. From this formal definition, much work has been completed by Bressloff, in extending snapping out BM to an applicable theory in developing a renewal description using the solutions of the radiation BVP \cite{bressloff2022probabilistic,bressloff2023renewal,bressloff2023renewal2}, as well as with applications such as stochastic resetting, multilayered media, narrow capture and entropy production \cite{bressloff2022probabilistic,bressloff2023renewal2,bressloff20233d,bressloff2024entropy}. 
In this letter we focus on a one-dimensional scenario and show that snapping out BM is not the full picture and that the permeable barrier problem can be reformulated in terms of the reflected BM (RBM) with the crossing process governed by a subordinated Poisson process. This provides a significant advance as RBM problems are notoriously easier to solve (i.e. its Green's function can be easily found via the method of images \cite{redner2001guide} or the defect technique \cite{kenkre2021memory}) as compared to the more complex BVP with radiation boundary conditions.

\section{Probabilistic Construction of Bronwian Motion in the Presence of Permeable Barriers}

\subsection{Subordination Procedure}

Building upon the snapping out BM \cite{lejay2006constructions,aho2016diffusion,bressloff2022probabilistic,bressloff2023renewal,bressloff2023renewal2}, we consider one-dimensional BM in an infinite domain in the presence of a permeable barrier at the origin through two interconnected stochastic processes, RBM, $Z(t)$, and the boundary local time, $\ell(t)$, of RBM, which is a stochastic quantity that characterises the amount of time the RBM spends at the reflecting barrier. Whenever the RBM is at the permeable barrier and the corresponding boundary local time $\ell(t)$ has exceeded the value of a random variable (RV) drawn from an exponential distribution (with mean $1/\kappa$ where $\kappa$ is the permeability), the RBM is allowed to pass through the barrier. Subsequently $\ell(t)$ is set to zero and RBM is now occurring on the other side of the barrier. A new exponential RV is drawn, and once $\ell(t)$  exceeds the RV, the RBM is allowed to pass back through the barrier and the process is repeated. 

\begin{figure}
    \centering \includegraphics[width=0.47\textwidth]{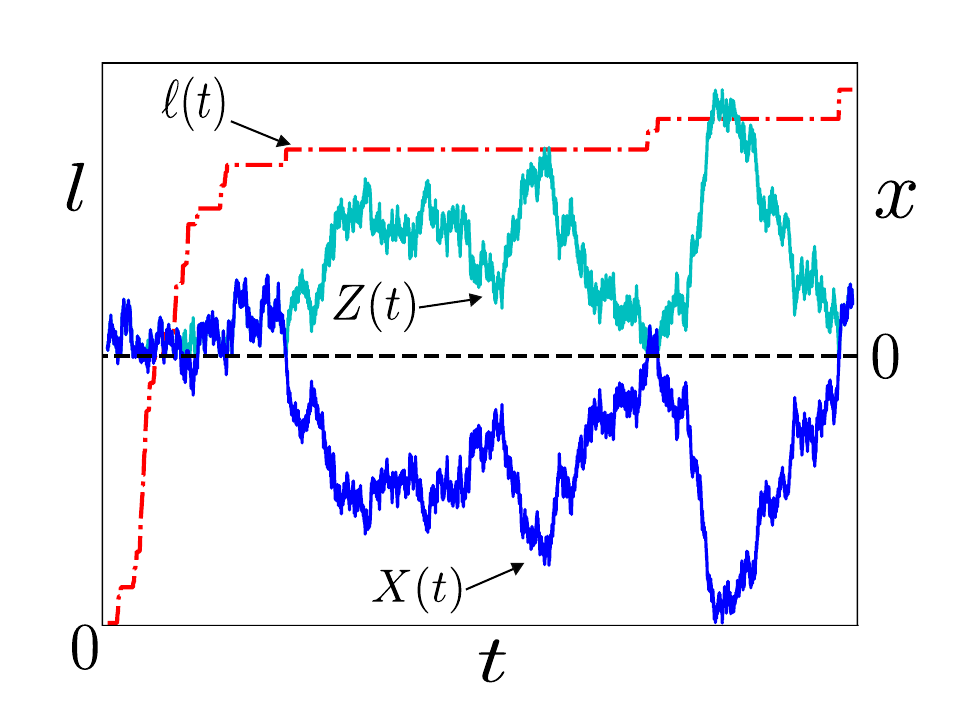}
    \caption{Position (right vertical axis) of a sample BM trajectory in the presence of a permeable barrier at the origin, $X(t)$, and of RBM, $Z(t)$, reflected at the origin. The local time of RBM at the origin, $\ell(t)$, is also plotted (left vertical axis) and acts as the subordinator of the Poisson process, which determines the crossing of the permeable barrier, see Eq. (\ref{eq:stochastic_representation}). Note that while $\ell(t)$ has dimensions $[\text{time}]/[\text{length}]$, $X(t)$ and $Z(t)$ have dimensions of $[\text{length}]$. 
    \label{fig:perm_local_time}}
\end{figure}

Due to the symmetry of RBM around the barrier, the total boundary local time on both sides of the permeable barrier together is equivalent to the boundary local time of RBM, see Fig. (\ref{fig:perm_local_time}). As the movement through the barrier occurs when $\ell(t)$ exceeds an exponential RV, the crossing process is a renewal process with exponential waiting times, that is a Poisson point process. However, the waiting times do not depend on the physical time, but rather the boundary local time. In other words the Poisson point process of crossing is subordinated to the stochastic process of $\ell(t)$, which acts as a stochastic clock. 

The above construction allows us to represent the location of a Brownian particle, $X(t)$, in the presence of a permeable barrier at the origin via
\begin{equation}
\label{eq:stochastic_representation}
    X(t)=(-1)^{N(\ell(t))}\big|x_0+\sqrt{2D}W(t)\big|.
\end{equation}
In Eq. (\ref{eq:stochastic_representation}) $D$ is the diffusion constant, $W(t)$ is the Wiener process, and the reflected Brownian motion (RBM) is such that $Z(t)=|x_0+\sqrt{2D}W(t)|$ with $X(0)=Z(0)=|x_0|$ (see Ref. \footnote{One could also represent the RBM in terms of the integral of the Skorokhod equation \cite{skorokhod1961stochastic,skorokhod1962stochastic,freidlin1985functional,saisho1987stochastic,ito1996diffusion,grebenkov2019probability}, $dZ(t)=\sqrt{2D}dW(t)+Dd\ell(t)$.} for an alternative representation of RBM). $N(l)$ represents a Poisson point process whose probability is given by the Poisson distribution, $\mathbb{P}[N(l)=n]=p_n(l)=(\kappa l)^n e^{-\kappa \l}/n!$, where $\kappa$ is the Poisson `rate' (units of $\kappa$ and $l$ are, respectively, [length]/[time] and [time]/[length]). The notation $N(\ell(t))$ indicates that the Poisson process is subordinated, i.e. undergone a stochastic time-change, to the  boundary local time of RBM at the origin, defined as \cite{levy1940certains,ito1996diffusion,mckean1975brownian,grebenkov2019probability}, 
\begin{equation}\label{eq:local_time}
    \ell(t)=\int_0^t\delta(Z(t'))dt'.
\end{equation}
From the above definition it is straightforward to verify that $\ell(t)$ meets all the requirements to be a subordinator (see e.g. Refs. \cite{applebaum2009levy,borodin2017stochastic}).

To prove that Eq. (\ref{eq:stochastic_representation}) is the correct representation of BM in the presence of a permeable barrier, we calculate the associated probability density of $X(t)$, $P(x,t|x_0)$. We proceed by first finding the joint density of $N(\ell(t))$ and $Z(t)$, $\mathdutchcal{P}_n(x,t|x_0)$. Using the properties of the Dirac-$\delta$ function one may write $\mathdutchcal{P}_n(x,t|x_0)$ 
$=\int_0^\infty \left\langle \delta_{N(l),n} \ \delta(Z(t)-x) \delta(\ell(t)-l) \right\rangle_{x_0} dl$
where $\delta_{m,n}$ is the Kronecker-$\delta$ and the angled brackets indicate an expectation conditioned on $Z(0)=|x_0|$ and $n_0=0$ (subscript $n_0$ omitted to lighten notation). Utilizing the independence between the Poisson process and RBM we are able to write,
\begin{equation}\label{eq:subordination}
\mathdutchcal{P}_n(x,t|x_0)=\int_0^\infty p_n(l)\rho(x,l,t|x_0)dl,
\end{equation}
where $\rho(x,l,t|x_0)$ is the joint density of $Z(t)$ and $\ell(t)$. To find $P(x,t|x_0)$ one needs to count how many times the barrier is crossed. Without loss of generality, we take that initially the process starts in $\mathbb{R}^+$. In such a case to be in $\mathbb{R}^+$ at time $t$ the barrier must have been crossed an even number of times, such that $N(\ell(t))$ is even, and conversely, for the process to be in $\mathbb{R}^-$ at time $t$, $N(\ell(t))$ must be odd. Summing over all even or odd crossing events we obtain
\begin{equation}\label{eq:even_odd}
    P(x,t|x_0)=\left\{
        \begin{array}{ll}
        \displaystyle \sum_{n \ \text{even}}\limits \mathdutchcal{P}_n(x,t|x_0) ,\ x \in \mathbb{R}^+, \\[10pt]
        \displaystyle \sum_{n \ \text{odd}}\limits \mathdutchcal{P}_n(x,t|x_0) ,\ x \in \mathbb{R}^-.
        \end{array}
        \right .
\end{equation} 
After finding $\rho(x,l,t|x_0)$, inserting it into Eq. (\ref{eq:subordination}) and performing the summations in Eq. (\ref{eq:even_odd}) (see Appendix \ref{appen:perm_sol}) one recovers $P(x,t|x_0)$ as derived by other means, e.g. in Refs. \cite{kosztolowicz2001random,grebenkov2014exploring2,kay2022diffusion}.

\subsection{Crossing Statistics}

\begin{figure}[htp]
    \centering
    \includegraphics[width=0.47\textwidth]{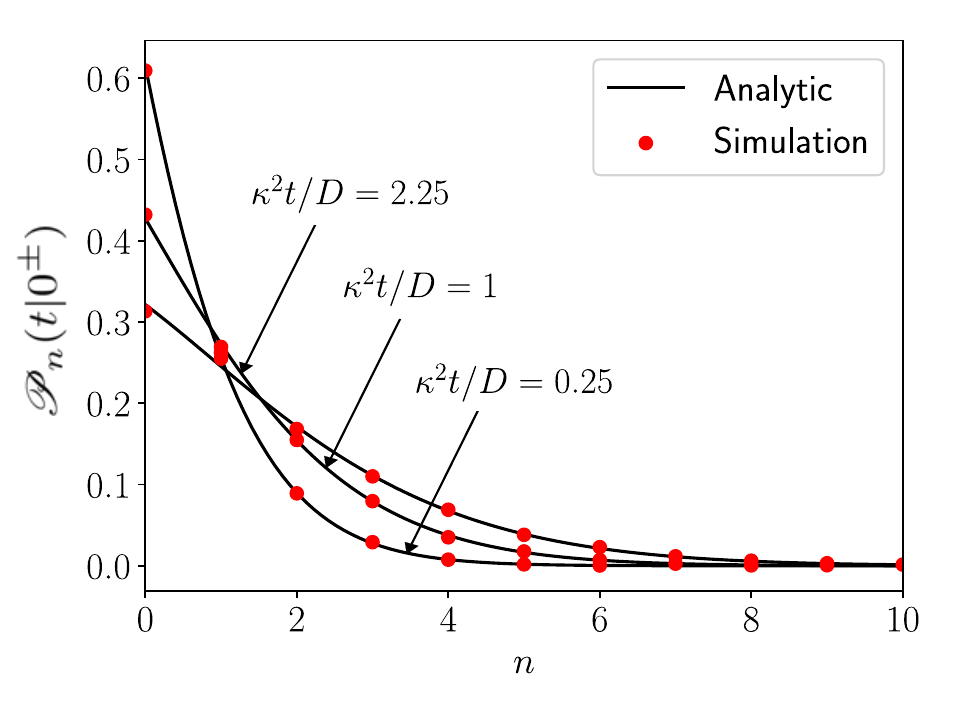}
    \caption{Plot of the number of crossings, $\mathdutchcal{P}_n(t|x_0)$, of a permeable barrier at the origin for a Brownian particle starting directly adjacent to the barrier. $\mathdutchcal{P}_n(t|x_0)$ is plotted using Eq. (\ref{eq:crossing_density}) and is compared to stochastic simulations for different values of the dimensionless parameter, $\kappa^2 t/D$. The dots are generated by simulating the snapping out BM \cite{lejay2016snapping} and counting the number of crossing events.}
    \label{fig:crossing}
\end{figure}

The subordination procedure in Eq. (\ref{eq:subordination}) allows one to investigate the number of times the barrier is crossed up to any time $t$, by simply marginalizing over $x$, i.e. $\mathdutchcal{P}_n(t|x_0)=\int_{0}^{\infty}p_n(l)\rho(l,t|x_0)$, where $\rho(l,t|x_0)=\int_0^{\infty}dx\,\rho(x,l,t|x_0)$. This quantity can be calculated (see Refs. \cite{takacs1995local,grebenkov2019probability}),  e.g. when $x_0=0$ it is given by $\rho(l,t|0)=\sqrt{D/\pi t} e^{-D l^2/4t}$. Multiplying by $p_n(l)$ and integrating over $l$ gives, $\mathdutchcal{P}_n(t|0^\pm)=f(n,\kappa^2t/D)$, where 
\begin{equation}\label{eq:crossing_density}
    f(n,y)=\pi^{-1/2}y^{n/2}U\left(\tfrac{n+1}{2},\tfrac{1}{2},y\right).
\end{equation}
$U(a,b,y)$ is Tricomi's confluent hypergeometric function \cite{bateman1953higher2}, and we have written $x_0=0^\pm$ to indicate directly to the right or left of the barrier, respectively, which are equivalent due to symmetry. This calculation can also be performed for $x_0\neq0$, but due to the cumbersome nature of the expression it is presented in Appendix \ref{appen:crossing_stat}. To confirm the validity of our theoretical development, we have compared Eq. (\ref{eq:crossing_density}) with stochastic simulations in Fig. \ref{fig:crossing}, showing perfect agreement.

The use of crossing statistics also provides a precise way of defining the permeability of the barrier, $\kappa$. Taking the average number of crossings, $\langle N(\ell(t))\rangle_{x_0}=\sum_{n=0}^\infty n \mathdutchcal{P}_n(t|x_0)$, using Eq. (\ref{eq:subordination}) with $\langle N(l) \rangle= \kappa l$, leads to the following identity for the permeability,
\begin{equation}\label{eq:perm_def}
    \kappa=\frac{\langle N(\ell(t))\rangle_{x_0}}{\langle \ell(t) \rangle_{x_0}}.
\end{equation}
Equation (\ref{eq:perm_def}) provides a theoretically grounded experimentally measurable definition of the permeability of a barrier. This intuitive expression can be understood and measured as the ratio of the number of times the barrier is crossed and the number of times the particle is reflected. 

\subsection{Governing Probability Equations}

The subordination procedure in Eq. (\ref{eq:subordination}) along with the method of calculating $P(x,t|x_0)$ in Eq. (\ref{eq:even_odd}) can be extended to multiple barriers, but it becomes a complicated combinatorial problem to find analytically the local time at multiple reflecting barriers. To bypass this difficulty we use Eqs. (\ref{eq:subordination}) and (\ref{eq:even_odd}) to find a set of governing equations which can be readily extended to an arbitrary number of barriers.

From Eq. (\ref{eq:subordination}), we take the time derivative of both sides, giving $\partial_t \mathdutchcal{P}_n(x,t|x_0)=\int_0^\infty p_n(l) \partial_t \rho(x,l,t|x_0)dl$. To find $\partial_t \rho(x,l,t|x_0)$ we take the inverse Laplace transform ($\alpha\rightarrow l$) of the Feynman-Kac equation, Eq. (\ref{eq:forward_FK}), which gives \cite{majumdar2007brownian}
\begin{equation}\label{eq:for_fk}
    \partial_t \rho(x,l,t|x_0)=[D\partial_x^2 -\delta(x)(\partial_l-\delta(l))]\rho(x,l,t|x_0).
\end{equation}
After inserting into the integral, and performing the integration, we get
\begin{align}
    \partial_t \mathdutchcal{P}_n(x,t|x_0)&=D \partial_x^2 \mathdutchcal{P}_n(x,t|x_0)+ p_n(0)\delta(x) \rho(x,0,t|x_0) \nonumber \\
    &-\delta(x)\int_0^\infty p_n(l) \partial_l \rho(x,l,t|x_0)dl.
\end{align}
Integrating the final term on the right-hand side by parts, gives
\begin{align}
    \partial_t \mathdutchcal{P}_n(x,t|x_0)&=D \partial_x^2 \mathdutchcal{P}_n(x,t|x_0)\nonumber \\
    &+\delta(x) \int_0^\infty dl\, \dot{p}_n(l) \rho(x,l,t|x_0),\label{eq:fk_deriv}
\end{align}
where $\dot{p}_n(l)= \frac{d}{dl} p_n(l)$ and $\lim_{l\to \infty} p_n(l)\rho(x,l,t|x_0)=0$. To deal with the last term in Eq. (\ref{eq:fk_deriv}) we use the fact that the Poisson distribution is governed by the following differential-difference equation \cite{gardiner1985handbook}, 
\begin{equation}\label{eq:poisson_ode}
    \dot{p}_n(l)=-\kappa p_n(l)+\kappa p_{n-1}(l).
\end{equation}
Inserting Eq. (\ref{eq:poisson_ode}) into Eq. (\ref{eq:fk_deriv}) and summing over even and odd $n$ and dropping $x_0$ from the notation we obtain
\begin{align}
    \partial_t P_+(x,t)&=D \partial_x^2 P_+(x,t) -\kappa \delta(x) \left[ P_+(x,t)-P_-(x,t)\right],\label{eq:birth_death1}\\
    \partial_t P_-(x,t)&=D \partial_x^2 P_-(x,t) -\kappa \delta(x) \left[ P_-(x,t)-P_+(x,t)\right]\label{eq:birth_death2},
\end{align}
with $\partial_x P_\pm(0,t)=0$, and where the $\pm$ subscript represents $P(x,t)$ valid for $x \in \mathbb{R}^\pm$, respectively. Equations (\ref{eq:birth_death1}) and (\ref{eq:birth_death2}) shall be termed birth-death diffusion equations, as one has source and sink terms localised at the origin  representing the passing through the barrier by entering and leaving the region. By integrating Eqs. (\ref{eq:birth_death1}) and (\ref{eq:birth_death2}) over the region $x\in[-\varepsilon,\varepsilon]$, taking $\varepsilon\to 0$ and utilizing $P_\pm(0,t)=P(0^\pm,t)$ it is easy to verify that the permeable barrier condition, $-D\partial_x P(0^\pm,t)=\kappa[P(0^-,t)-P(0^+,t)]$, is satisfied and that $P(x,t)=P_+(x,t)\Theta(x)+P_-(x,t)\Theta(-x)$ satisfy the diffusion equation away from the permeable boundary, for $\Theta(z)$ the Heaviside step function. Alternatively, one can show that $P(x,t)$ satisfies the equation $\partial_t P(x,t)=D\partial_x^2 P(x,t)+(D/\kappa)\delta^{\prime}(x)J(0,t)$ with the flux $J(0,t)=-D\partial_x P(0,t)$ \cite{kay2022diffusion,kay2023extreme}.

To solve Eqs. (\ref{eq:birth_death1}) and (\ref{eq:birth_death2}) one can translate these equations into renewal type equations, in terms of the Green's function of RBM, $G(x,t|x_0)$ (i.e. $G(x,0|x_0)=\delta(x-x_0)$ and $\partial_x G(0,t|x_0)=0)$, for $x_0\in \mathbb{R}^+$,
\begin{align}
    P_+(x,t|x_0)&=G(x,t|x_0)
    -\kappa \int_0^t ds\, G(x,t-s|0) \nonumber \\
    &\times\left[P_+(0,s|x_0)-P_-(0,s|x_0)\right],\label{eq:birth_death_renewal1}\\
    P_-(x,t|x_0)&=-\kappa \int_0^t ds\, G(x,t-s|0) \nonumber \\
    &\times\left[P_-(0,s|x_0)-P_+(0,s|x_0)\right].\label{eq:birth_death_renewal2}
\end{align}
Equations (\ref{eq:birth_death_renewal1}) and (\ref{eq:birth_death_renewal2}) can be solved by utilizing the convolution nature of the integrals. After Laplace transforming (i.e. $\widetilde{f}(\epsilon)=\int_0^\infty e^{-\epsilon t}f(t)dt$), setting $x=0$ and rearranging (see e.g. Refs. \cite{kenkre2021memory,kay2022defect}), one has the following solutions,
\begin{align}
    \widetilde{P}_+(x,\epsilon|x_0)&=\widetilde{G}(x,\epsilon|x_0)-\widetilde{G}(x,\epsilon|0)\frac{\widetilde{G}(0,\epsilon|x_0)}{\frac{1}{\kappa}+2 \widetilde{G}(0,\epsilon|0)}, \label{eq:birth_death_renewal_sol1}\\
    \widetilde{P}_-(x,\epsilon|x_0)&=\widetilde{G}(x,\epsilon|0)\frac{\widetilde{G}(0,\epsilon|x_0)}{\frac{1}{\kappa}+2 \widetilde{G}(0,\epsilon|0)} \label{eq:birth_death_renewal_sol2}.
\end{align}
With $\widetilde{G}(x,\epsilon|x_0)\!=\!\left(\!e^{-|x-x_0|\sqrt{\!\epsilon/D}}\!+\!e^{-|x+x_0|\sqrt{\!\epsilon/D}}\right)\!/\!\sqrt{2D\epsilon}$ inserted into Eqs. (\ref{eq:birth_death_renewal_sol1}) and (\ref{eq:birth_death_renewal_sol2}) one obtains the known solution as in Ref. \cite{kay2022diffusion}. Eqs. (\ref{eq:birth_death_renewal1}) and (\ref{eq:birth_death_renewal2}) are still valid if there are external forces present, where one would instead need the (Laplace transformed) Green's function of the corresponding Smoluchowski equation which can be inserted into Eqs. (\ref{eq:birth_death_renewal_sol1}) and (\ref{eq:birth_death_renewal_sol2}) to give the required solution. 

\section{Application to Asymmetric Barriers}

An important generalization is the extension of Eqs. (\ref{eq:birth_death1}) and (\ref{eq:birth_death2}) to when the barrier is asymmetric, i.e.
$-D\partial_x P(0^\pm,t)=\kappa_-P(0^-,t)-\kappa_+P(0^+,t)$ 
\cite{kosztolowicz2021boundary,bressloff2023asymmetric}, which reduces to the radiation/Robin boundary condition associated with elastic BM when either $\kappa_+$ or $\kappa_-$ vanishes. In the context of Eq. (\ref{eq:stochastic_representation}) $\kappa_+\not=\kappa_-$ corresponds to having a different Poisson rate for even and odd crossings of the barrier. To satisfy the asymmetric barrier condition, one modifies Eqs. (\ref{eq:birth_death1}) and (\ref{eq:birth_death2}) as follows,
\begin{align}
    \partial_t P_+(x,t)&=D \partial_x^2 P_+(x,t) \nonumber \\
    &-\delta(x) \left[ \kappa_+ P_+(x,t)-\kappa_- P_-(x,t)\right],\label{eq:birth_death_asym_1}\\
    \partial_t P_-(x,t)&=D \partial_x^2 P_-(x,t) \nonumber\\
    &-\delta(x) \left[ \kappa_- P_-(x,t)-\kappa_+ P_+(x,t)\right]\label{eq:birth_death_asym_2},
\end{align}
with the reflecting BC at the origin. Equations (\ref{eq:birth_death_asym_1}) and (\ref{eq:birth_death_asym_2}) admit renewal equations akin to Eqs. (\ref{eq:birth_death_renewal1}) and (\ref{eq:birth_death_renewal2}), and thus lead to the closed form solutions in terms of the Green's functions of RBM (for $x_0\in \mathbb{R}^+$),
\begin{align}
    \widetilde{P}_+(x,\epsilon|x_0)&=\widetilde{G}(x,\epsilon|x_0)-
     \frac{\kappa_+ \widetilde{G}(x,\epsilon|0) \widetilde{G}(0,\epsilon|x_0)}{1+(\kappa_++\kappa_-)\widetilde{G}(0,\epsilon|0)},\\
    \widetilde{P}_-(x,\epsilon|x_0)&= \frac{\kappa_+ \widetilde{G}(x,\epsilon|0) \widetilde{G}(0,\epsilon|x_0)}{1+(\kappa_++\kappa_-)\widetilde{G}(0,\epsilon|0)},
\end{align}
and the analogous solution when $x_0\in \mathbb{R}^-$,
\begin{align}
    \widetilde{P}_+(x,\epsilon|x_0)&=
     \frac{\kappa_- \widetilde{G}(x,\epsilon|0) \widetilde{G}(0,\epsilon|x_0)}{1+(\kappa_++\kappa_-)\widetilde{G}(0,\epsilon|0)},\\
    \widetilde{P}_-(x,\epsilon|x_0)&= \widetilde{G}(x,\epsilon|x_0)-\frac{\kappa_- \widetilde{G}(x,\epsilon|0) \widetilde{G}(0,\epsilon|x_0)}{1+(\kappa_++\kappa_-)\widetilde{G}(0,\epsilon|0)}.
\end{align}
Extension of the birth-death equation for asymmetric barriers to higher dimensions, and their solution in the Laplace domain, is presented in Appendix \ref{appen:higer_dimensions}.

A statistical representation of the asymmetric permeabilities, $\kappa_-$ and $\kappa_+$, akin to Eq. (\ref{eq:perm_def}), can also be defined. The distinction arises whereby for the asymmetric case one would measure the mean number of crossing events from one direction and the corresponding mean local time to determine, by taking the ratio, the specific permeability for one side of the barrier.

\subsection{Arrays of Barriers}

\begin{figure}[h!]
    \centering
    \includegraphics[width=0.47\textwidth]{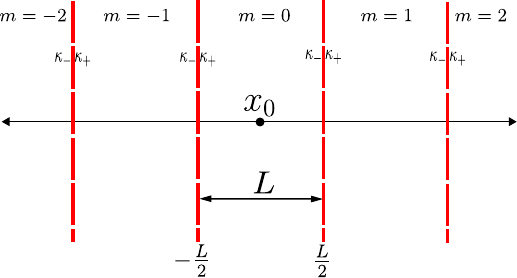}
    \caption{Diagram showing the arrangement of periodically placed identical asymmetric barriers, separated from each other by a distance $L$, with permeability $\kappa_-$ and $\kappa_+$ for the left and right-hand side of the barrier, respectively.}
    \label{fig:perm_array}
\end{figure}

Here we utilize Eqs. (\ref{eq:birth_death_asym_1}) and (\ref{eq:birth_death_asym_2}) to study a system of BM in the presence of an infinite number of periodically placed identical asymmetric permeable barriers. We place the asymmetric permeable barriers at $x=L/2+m L$, where we denote each compartment as $m\in \mathbb{Z}$ with $x_0$ being in the $m=0$ compartment, see Fig. (\ref{fig:perm_array}).

We extend Eqs. (\ref{eq:birth_death_asym_1}) and (\ref{eq:birth_death_asym_2}) to the situation displayed in Fig. (\ref{fig:perm_array}). The purpose of this extension is to explore and quantify directed motion emerging as a result of the asymmetric nature of the barrier permeability, a process akin to the so called noise-induced drift in the presence of thermal ratchet potentials \cite{magnasco1993forced,astumian1994fluctuation}. Labelling the probability density for a single compartment $m$ as $P_m(x,t|x_0)$ (see Fig. (\ref{fig:perm_array}), where $x\in [-L/2+mL,L/2+mL]$, we arrive at the following,

\begin{align}
\partial_t &P_m(x,t)=D\partial^2_x P_m(x,t)-\delta\left(x-\tfrac{(2m+1)L}{2}\right) \nonumber \\
     &\times\big[\kappa_- P_m(x,t) - \kappa_+ P_{m+1}(x,t)\big] \nonumber \\
     &\!-\!\delta\left(x-\tfrac{(2m-1)L}{2}\right) 
     \big[\kappa_+P_m(x,t) \!-\!\kappa_-P_{m-1}(x,t)  \big]\label{eq:array_birth_death},
\end{align}
with the BCs, $\lim_{x\to \pm L/2+mL} \partial_x P_m(x,t|x_0)=0$.
As before we may write the Laplace transformed solution of Eq. (\ref{eq:array_birth_death}) in terms of the associated Green's functions, which satisfies the reflecting BCs $\lim_{x\to \pm L/2+mL} \partial_x G(x,t)=0$. Therefore, in the Laplace domain we have the representation,  \begin{align}\label{eq:array_sol}
&\widetilde{P}_m(x,\epsilon|x_0)=\widetilde{G}(x,\epsilon|x_0)\delta_{m,0}-\widetilde{G}\left(x,\epsilon\Big|\tfrac{(2m+1)L}{2}\right)\nonumber \\ &\times\left[\kappa_-\widetilde{P}_m\left(\tfrac{(2m+1)L}{2},\epsilon\Big|x_0\right)-\kappa_+ \widetilde{P}_{m+1}\left(\tfrac{(2m+1)L}{2},\epsilon\Big|x_0\right)\right]
    \nonumber \\
    &-\widetilde{G}\left(x,\epsilon\Big|\tfrac{(2m-1)L}{2}\right)\left[\kappa_+ \widetilde{P}_m\left(\tfrac{(2m-1)L}{2},\epsilon\Big|x_0\right)\right. \nonumber \\
    &\qquad\left.-\kappa_- \widetilde{P}_{m-1}\left(\tfrac{(2m-1)L}{2},\epsilon\Big|x_0\right)\right],
\end{align}

As the Green's function $\widetilde{G}(x,\epsilon|x_0)$ is known, we may solve Eq. (\ref{eq:array_sol}). To do so, we set $x=(2m+1)L/2$ and $x=(2m-1)L/2$ to form two simultaneous equations. A careful look at the form of the Green's function shows that for these values of $x$ the $m$ dependence drops out. One is then able to exploit the recurrence relation and utilize discrete Fourier transforms to find the exact form of $\widetilde{P}_m(x,\epsilon|x_0)$ (see Appendix \ref{appen:inf_array}).

We extract effective transport parameters that describe the dynamics at long-times by calculating the first and second moment of $\widetilde{P}_m(x,\epsilon|x_0)$ (see Appendix \ref{appen:inf_array}). We find that the mean, $\langle X(t) \rangle$ in the long-time limit, $t\to \infty$ leads to, $\langle X(t)\rangle \sim \nu_\text{eff} \ t$, up to leading order, where $\nu_\text{eff}$ is the effective velocity or drift, and is given by
\\
\begin{equation}\label{eq:eff_vel}
    \nu_\text{eff}=\frac{2D(\kappa_--\kappa_+)}{2D+L(\kappa_-+\kappa_+)}.
\end{equation}

Equation (\ref{eq:eff_vel}) shows that the direction of the effective velocity is determined by the relative magnitudes of $\kappa_+$ and $\kappa_-$, such that its direction is along the path of least resistance from the barriers (higher permeability). The distance between the barriers also affects $\nu_\text{eff}$, where a smaller $L$ leads to more interactions with the barriers and less space to diffuse freely, causing an increase in $\nu_\text{eff}$. 

To find the particle dispersion we construct the second moment, $\langle X^2(t) \rangle$, which scales as $\langle X^2(t) \rangle \sim 2 D_\text{eff} t + \nu_\text{eff}^2 t^2$ for $t\to \infty$, with an effective diffusion constant,
\begin{align}
\label{eq:diff_eff}
    D_\text{eff}\!&=\!\frac{D L}{6} \Big[24 D^2 (\kappa_-\!+\!\kappa_+)+2 D L \left(7 \kappa_-^2\!+\!34 \kappa_- \kappa_+\!+\!7 \kappa_+^2\right) \nonumber \\
    &+L^2 (\kappa_-+\kappa_+) \left(5 \kappa_-^2+14 \kappa_- \kappa_++5 \kappa_+^2\right)\Big] \nonumber \\
    &\times\Big[2 D+L (\kappa_-+\kappa_+)\Big]^{-3}.
\end{align}

When the permeability is symmetric ($\kappa_+=\kappa_-=\kappa$), $\nu_\text{eff}=0$ and any source of directional bias disappears, while the effective diffusion constant reduces to the known form, $D_\text{eff}=DL\kappa/(D+L\kappa)$ \cite{powles1992exact,dudko2004diffusion,kenkre2008molecular}. 

\subsection{Non-Equilibrium Steady-State}

Finally, we consider the related problem of an asymmetric permeable barrier in a periodic domain. This scenario displays the same effect of an infinite array of periodic barriers, but with the addition of the emergence of a steady state in the long time limit, due to the finiteness of the domain. If we place for simplicity an asymmetric permeable barrier at the origin in a domain of $x\in[-L/2,L/2]$, with periodic boundaries, the steady state is found by solving the stationary diffusion equation, $P''(x)=0$ with the following conditions, $P(-L/2)=P(L/2), -D P'(0^-)=-DP'(0^+)=\kappa_- P(0^-)-\kappa_+ P(0^+), \int_{-L/2}^{L/2} P(x)dx=1$, this simple calculation leads to, 
\begin{equation}\label{eq:ness}
    P(x)=\left\{
    \begin{array}{ll}
         & \displaystyle\frac{2 [D + L \kappa_+ + x (\kappa_+-\kappa_-)]}{L[2D+L(\kappa_-+\kappa_+)]}, \ x \in [-L/2,0), \\[10pt]
         & \displaystyle\frac{2 [D + L \kappa_- + x (\kappa_+-\kappa_-)]}{L[2D+L(\kappa_-+\kappa_+)]}, \ x \in (0,L/2].
    \end{array}
    \right .
\end{equation}
The associated non-vanishing flux from Eq. (\ref{eq:ness}), given by $J(x)=-D P'(x)=\nu_\text{eff}/L$,  shows the existence of a non-equilibrium steady state (NESS), which disappears when $\kappa_-=\kappa_+$. In other words, the presence of an asymmetric barrier in a periodic domain, keeps the system out of equilibrium by imposing an effective bias throughout the system.

\section{Conclusion}

To summarize, we have developed a representation alternative to refs. \cite{bressloff2023asymmetric} of the spatio-temporal dynamics of a diffusive  particle in the presence of an asymmetric thin permeable barrier dynamics. Our approach makes use of a stochastic description of Brownian motion (BM) with permeable barriers, using reflected Brownian motion (RBM) and its local time at the barrier which acts as a subordinator of a Poisson process for the barrier crossing events. This has led to an analytic representation of the probability distribution of the number of crossing and an empirically relevant definition of barrier permeability as the ratio of mean crossings to mean local time. Applying Feynman-Kac theory, we have derived macroscopic governing equations for BM with permeable barriers, resulting in coupled birth-death diffusion equations and their solutions in the Laplace domain, and extended to asymmetric barriers. Finally, we have considered a Brownian particle diffusing through an infinite array of periodically placed identical asymmetric barriers as well as through a single asymmetric permeable barrier in a spatially periodic domain. For the unbounded system by deriving the exact probability distribution in space and Laplace domain, we have been able to quantify the appearance of an effective velocity at long times. In doing so we have uncovered a noise-induced drift phenomenon for a diffusing particle without invoking external forces or modifying motion characteristics of the Brownian particle (compare e.g. ref. \cite{robertszhen2023} where a ratchet potential is subject to dichotomous noise leading to alternative types of coupled occupation probability in each compartment). For the finite periodic domain we have shown explicitly the appearance of a NESS.


The present work may be further developed in various directions. One consists of having the barrier crossing process no longer Poissonian but governed by an arbitrary renewal process. Extensions of practical value to empirical observations consists of modifying the underlying motion through external potentials or through changes in the movement statistics from diffusive to subdiffusive, for which a closely related radiation boundary condition has been proposed \cite{kay2023subdiffusion}. Starting from our exact solution of the infinite array of periodically placed asymmetric permeable barriers, another fruitful direction is to develop an effective medium approximation \cite{parris2008random,kenkre2009extensions} for when the positions and/or the strengths of the periodic array of barriers are perturbed. Such study would allow to identify under what conditions hop-diffusion becomes anomalous \cite{slkezak2021diffusion}, a topic of direct relevance to the motion of biomolecules in the plasma membrane of eukaryotic cells \cite{ritchie2005detection,fujiwara2016confined,krapf2018compartmentalization}. Finally, the exact analytic findings for the finite domain with NESS could be exploited  to study thermodynamics fluctuations relations using techniques from large-deviation theory \cite{dieball2023direct}.

\acknowledgements

TK and LG would like to thank Hernan Larralde for useful discussions. TK and LG acknowledge funding from, respectively, an Engineering and Physical Sciences Research
Council (EPSRC) EP/T517872/1, and the Natural and Natural Environment Research Council (NERC) Grant No. NE/W00545X/1.

\appendix

\section{Calculation of the Permeable Barrier Solution from Poisson Formulation}\label{appen:perm_sol}

Here we show how the solution of the DE with the symmetric permeable barrier condition, $-D \partial_x P(0,t)=\kappa\left[P(0^-,t)-P(0^+,t)\right]$ is recovered from the sums in Eq. (\ref{eq:even_odd}) of the main text. Recalling the definition of $\mathdutchbcal{P}_n(x,t|x_0)$ in Eq.  (\ref{eq:subordination}) of the main text, and carrying the sums over $p_n(l)$, Eq. (\ref{eq:even_odd}) becomes,
\begin{equation}\label{eq:even_odd_sub_permeable}
    P(x,t|x_0)=\left\{
        \begin{array}{ll}
        \displaystyle \int_0^\infty e^{-\kappa l} \cosh(\kappa l) \rho(x,l,t|x_0)dl ,\ x \in \mathbb{R}^+, \\[20pt]
        \displaystyle \int_0^\infty e^{-\kappa l} \sinh(\kappa l) \rho(x,l,t|x_0)dl ,\ x \in \mathbb{R}^-.
        \end{array}
        \right .
\end{equation} 
All one needs is the the solution of the (forward) Feynman-Kac equation, $\varrho(x,\alpha,t|x_0)$, where $\rho(x,l,t|x_0)=\int_0^\infty e^{-\alpha l}\varrho(x,\alpha,t|x_0)dl$, such that
\begin{equation}\label{eq:forward_FK}
    \frac{\partial}{\partial t} \varrho(x,\alpha,t|x_0)=\left[D\frac{\partial^2}{\partial x^2} -\alpha\delta(x)\right]\varrho(x,\alpha,t|x_0),
\end{equation}
with $\varrho(x,\alpha,0|x_0)=\delta(x-x_0)$ and $\partial_x \varrho(0,\alpha,t|x_0)=0$. The solution to Eq. (\ref{eq:forward_FK}) can be simply found from the Green's function of the diffusion equation with reflection at $x=0$, i.e. for $\alpha=0$ in Eq. (\ref{eq:forward_FK}), $\varrho(x,0,t|x_0)=G(x,t|x_0)$, leading to the following solution in the Laplace domain \cite{kenkre2021memory,kay2022defect}
\begin{equation}
    \widetilde{\varrho}(x,\alpha,\epsilon|x_0)=\widetilde{G}(x,\epsilon|x_0)-\widetilde{G}(x,\epsilon|0)\frac{\widetilde{G}(0,\epsilon|x_0)}{\frac{1}{\alpha}+\widetilde{G}(0,\epsilon|0)},
\end{equation}
where $\widetilde{G}(x,\!\epsilon|x_0)\!=\!\left(\!e^{-|x-x_0|\sqrt{\!\epsilon/D}}\!+\!e^{-|x+x_0|\sqrt{\!\epsilon/D}}\right)\!/\!\sqrt{2D\epsilon}$. After inverse Laplace transforming with respect to $\alpha\to l$ and $\epsilon\to t$, one obtains,
\begin{align}\label{eq:local_time_reflect_joint}
    \rho(x,l,t|x_0)&=\frac{e^{-\frac{(Dl+ | x| +| x_0| )^2}{4 D t}}}{\sqrt{4 \pi D t^3 } } (Dl+ | x| +| x_0| ) \nonumber \\
    &+ \delta (l) \frac{ e^{-\frac{| x-x_0| ^2}{4 D t}}-2 e^{-\frac{(| x| +| x_0| )^2}{4 D t}}+e^{-\frac{| x+x_0| ^2}{4 D t}}}{\sqrt{4 \pi D t} }.
\end{align}
\begin{widetext}
Inserting Eq. (\ref{eq:local_time_reflect_joint}) into Eq. (\ref{eq:even_odd_sub_permeable}) and performing the integrations one finds,
\begin{equation}\label{eq:perm_sol_sub}
    P(x,t|x_0)=\left\{
    \begin{array}{ll}
    \dfrac{e^{\frac{-(x+x_0)^2}{4 D t}}+e^{\frac{-(x-x_0)^2}{4 D t}}}{\sqrt{4 \pi D t}}\\
    -\dfrac{\kappa}{D}  e^{2 \kappa  (x+x_0 +2 \kappa  t)/D} \erfc\left(\frac{x+x_0 +4 \kappa  t}{2 \sqrt{D t}}\right),\ x \in \mathbb{R}^+, \\[20pt]
    \dfrac{\kappa}{D} e^{2 \kappa  (x_0-x +2 \kappa  t)/D} \erfc\left(\frac{x_0-x +4 \kappa  t}{2 \sqrt{D t}}\right),\ x \in \mathbb{R}^-,
    \end{array}
    \right .
\end{equation}
which is exactly the well known solution to the diffusion equation with a permeable barrier at the origin, e.g.\cite{powles1992exact,grebenkov2014exploring1,kay2022diffusion}.

\section{Crossing Statistics for a Brownian Particle starting at $x_0$}\label{appen:crossing_stat}

To calculate the distribution of the number of crossings up to time $t$, $\mathdutchcal{P}_n(t|x_0)$, for a Brownian particle starting at $X(0)=x_0$, we start from the definition $\mathdutchcal{P}_n(t|x_0)=\int_0^\infty p_n(l) \rho(l,t|x_0)dl$, where $\rho(l,t|x_0)$ is the marginal over $x$ of Eq. (\ref{eq:local_time_reflect_joint}). This marginalization leads to 
\begin{equation}
    \rho(l,t|x_0)=\sqrt{\frac{D}{\pi t}} e^{-\frac{(Dl+|x_0|)^2}{4Dt}}+\delta(l)\erf\left(\frac{|x_0|}{2\sqrt{D t}}\right).
\end{equation}
After integration over $l$ we find $\mathdutchcal{P}_n(t|x_0)$ to be 
\begin{multline}\label{eq:crossing_density_full}
    \mathdutchcal{P}_n(t|x_0)=\left(\tfrac{\kappa^2 t}{D}\right)^{n/2} e^{-\frac{x_0^2}{4 D t}} \Bigg[\frac{1}{\Gamma \left(\frac{n}{2}+1\right)}\, _1F_1\left(\frac{n+1}{2};\frac{1}{2};\frac{t}{D} \left(\frac{|x_0|}{2 t}+\kappa \right)^2\right)
    \\-\frac{(|x_0|+2 \kappa  t) }{\sqrt{D t} \Gamma \left(\frac{n+1}{2}\right)} \, _1F_1\left(\frac{n}{2}+1;\frac{3}{2};\frac{t}{D} \left(\frac{|x_0|}{2 t}+\kappa \right)^2\right)\Bigg]
    + \delta_{n,0} \erf\left( \frac{|x_0|}{2\sqrt{D t}}\right),
\end{multline}

where $_1F_1(a;b;z)$ is the confluent hypergeometric function \cite{bateman1953higher2}. Setting $x_0=0$ one recovers Eq. (\ref{eq:crossing_density}).

\section{Birth-Death Renewal Equations in Higher Dimensions}\label{appen:higer_dimensions}

The birth-death formalism in the main text can be directly extended to dimension $d>1$. We take an asymmetric permeable barrier to enclose the domain $\Omega \subset \mathbb{R}^d$, with the permeable barrier being a hypersurface that can be approached from $\Omega$ or $\mathbb{R}^d \setminus \Omega$ which is represented by $\partial \Omega^-$ and $\partial \Omega^+$, respectively. We are then able to write the higher dimensional renewal type equations corresponding to renewal version of Eqs. (15) 
and (16), 
for a permeable hypersurface
\begin{align}
    P_+(\mathbf{x},t|\mathbf{x}_0)&=G_+(\mathbf{x},t|\mathbf{x}_0)- \int_0^t ds \int_{\partial \Omega} d\mathbf{y} G_+(\mathbf{x},t-s|\mathbf{y})\left[\kappa_+ P_+(\mathbf{y},s|\mathbf{x}_0)-\kappa_- P_-(\mathbf{y},s|\mathbf{x}_0)\right] ,\label{eq:birth_death_renewal1_d}\\
    P_-(\mathbf{x},t|\mathbf{x}_0)&=-\int_0^t ds \int_{\partial \Omega} d\mathbf{y} G_-(\mathbf{x},t-s|\mathbf{y})\left[\kappa_-P_-(\mathbf{y},s|\mathbf{x}_0)-\kappa_+P_+(\mathbf{y},s|\mathbf{x}_0)\right],\label{eq:birth_death_renewal2_d}
\end{align}
for $\mathbf{x}_0 \in \mathbb{R}^d\setminus \Omega$. $G_+(\mathbf{x},t|\mathbf{x}_0)$ and $G_-(\mathbf{x},t|\mathbf{x}_0)$ represent the Green's function for RBM for $\mathbf{x} \in \mathbb{R}^d \setminus \Omega$ and $\mathbf{x} \in \Omega$, respectively. The reason we need two different Green's functions here compared to the case in the main text is that in $d=1$ we have symmetry about the barrier whereas in general for $d>1$ we do not. For certain geometries, such as the ones that display radial symmetry, the Green's functions are known (for e.g. see Refs. \cite{carslaw1962conduction,redner2001guide}) and one can solve the above renewal equation in a similar manner to that in the main text, we highlight this below.

Consider we have a particle diffusing in $\mathbf{x}\in\mathbb{R}^d$, and let us consider the situation in which we have radial symmetry, such that the position of the particle is fully defined by $r=|\mathbf{x}|$, and we have the symmetric initial condition of $\delta(r-r_0)/\Sigma_d(r_0)$, where $\Sigma_d(r_0)$ is the surface area of the hypersphere at radius $r_0$. Suppose we place a permeable barrier hypershere of radius $R$, and take $r_0>R$, then Eqs. (\ref{eq:birth_death_renewal1_d}) and (\ref{eq:birth_death_renewal2_d}) lead to,
\begin{align}
    P_+(r,t|x_0)&=G_+(r,t|r_0)-\Sigma_d(R) \int_0^t G_+(r,t-s|R)\left[\kappa_+ P_+(R,t|r_0)-\kappa_- P_-(R,t|r_0)\right],\label{eq:birth_death_renewal1_radial}\\
    P_-(r,t|x_0)&=-\Sigma_d(R) \int_0^t G_-(r,t-s|R)\left[\kappa_- P_-(R,t|r_0)-\kappa_+ P_+(R,t|r_0)\right],\label{eq:birth_death_renewal2_radial}
\end{align}
where the '$\pm$' subscripts represent $r\in[R^+,\infty)$ and $r \in [0,R^-]$. After Laplace transforming and rearranging, we find the solutions as,
\begin{align}
    \widetilde{P}_+(r,\epsilon|r_0)&=\widetilde{G}_+(r,\epsilon|r_0)-\frac{\kappa_+\widetilde{G}_+(r,\epsilon|R)\widetilde{G}_+(R,\epsilon|r_0)}{\frac{1}{\Sigma_d(R)}+\kappa_+\widetilde{G}_+(R,\epsilon|R)+\kappa_-\widetilde{G}_-(R,\epsilon|R)},\\
    \widetilde{P}_-(r,\epsilon|r_0)&=\frac{\kappa_+\widetilde{G}_-(r,\epsilon|R)\widetilde{G}_+(R,\epsilon|r_0)}{\frac{1}{\Sigma_d(R)}+\kappa_+\widetilde{G}_+(R,\epsilon|R)+\kappa_-\widetilde{G}_-(R,\epsilon|R)}.
\end{align}
Similarly this can be done for $r_0<R$. The Green's functions $G_\pm(r,t|x_0)$ for $d=2,3$ can be found for e.g. in Ref. \cite{carslaw1962conduction}.

One can see that the Eqs. (\ref{eq:birth_death_renewal1_radial}) and (\ref{eq:birth_death_renewal2_radial}) correspond to the following birth-death diffusion equations,
\begin{align}
    \partial_t P_+(r,t)&=D\left( \partial_r^2+\frac{d-1}{r}\partial_r \right) P_+(r,t)- \delta(r-R)\left[\kappa_+ P_+(r,t)-\kappa_-P_-(r,t) \right],\label{eq:birth_death1_radial}\\
    \partial_t P_-(r,t)&=D\left( \partial_r^2+\frac{d-1}{r}\partial_r \right) P_-(r,t)- \delta(r-R)\left[\kappa_- P_-(r,t)-\kappa_+ P_+(r,t) \right].\label{eq:birth_death2_radial}
\end{align}
By integrating over the Dirac-$\delta$ in Eqs. (\ref{eq:birth_death1_radial}) and (\ref{eq:birth_death2_radial}), we find the asymmetric permeable BC for radially symmetric systems,
\begin{equation}
    -D\partial_rP(R^\pm,t)=\left[\kappa_- P(R^-,t)- \kappa_+ P(R^+,t) \right],
\end{equation}
where $P(R^\pm,t)=P_\pm(R,t)$.

\section{Infinite Array of Identical Asymmetric Permeable Barriers}\label{appen:inf_array}

We  utilize the birth-death diffusion equations for an infinite array of asymmetric permeable barriers (see Fig. (\ref{fig:perm_array}) and Eq. (\ref{eq:array_birth_death}) in the main text) to find the exact propagator, $P_m(x,t)$, for any compartment $m$ in the array. For that we utilize the solution for the Laplace transform, $\widetilde{P}_m(x,\epsilon)$, in terms of the Green's functions $\widetilde{G}(x,\epsilon|x_0)$ in each domain. 

One proceeds by Laplace transforming the diffusion equation whose Green's functions obey, $\epsilon \widetilde{G}(x,\epsilon|x_0)-\delta(x-x_0)=D\partial^2_x\widetilde{G}(x,\epsilon|x_0)$, in the region $x,x_0 \in [-L/2+mL,L/2+mL]$ with the BCs $\lim_{x\to \pm L/2+mL}\partial_x \widetilde{G}(x,\epsilon|x_0)=0$. We then solve this equation for each of the following cases, $x<x_0$ and $x>x_0$, with the respective BCs, and ensuring continuity of $\widetilde{G}(x,\epsilon|x_0)$ at $x=x_0$, and after integrating over the Dirac-$\delta$ function, we also have the condition, $\partial_x \widetilde{G}(x_0^+,\epsilon|x_0)-\partial_x \widetilde{G}(x_0^-,\epsilon|x_0)=-D^{-1}$. After solving the differential equation and satisfying all of these conditions, one finds the Green's function as,
\begin{equation}\label{eq:two_ref_greens}
    \widetilde{G}(x,\epsilon|x_0)=\left\{
    \begin{array}{ll}
        \frac{\text{cosech}\left(\frac{L \sqrt{\epsilon }}{\sqrt{D}}\right) \left(\cosh \left(\frac{\sqrt{\epsilon } (2 L m-x-x_0)}{\sqrt{D}}\right)+\cosh \left(\frac{\sqrt{\epsilon } (L-x+x_0)}{\sqrt{D}}\right)\right)}{2 \sqrt{D} \sqrt{\epsilon }}, \ -L/2+mL\leq x_0\leq x\leq L/2+mL,
        \\[10pt]
        \frac{\text{cosech}\left(\frac{L \sqrt{\epsilon }}{\sqrt{D}}\right) \left(\cosh \left(\frac{\sqrt{\epsilon } (2 L m-x-x_0)}{\sqrt{D}}\right)+\cosh \left(\frac{\sqrt{\epsilon } (L+x-x_0)}{\sqrt{D}}\right)\right)}{2 \sqrt{D} \sqrt{\epsilon }}, \ -L/2+mL\leq x\leq x_0 \leq L/2+mL.
        
    \end{array}
    \right.
\end{equation}

\subsection{Solution within each compartment}

With the exact form of the Green's functions, we insert $x=\pm L/2+mL$ into Eq. (\ref{eq:array_sol}) to find the coupled equations   (for simplicity we set $x_0=0$ and we drop the symbol $x_0$ from $P_m((2m\pm 1)L/2,\epsilon|x_0)$ to lighten up notation),
\begin{align}\label{eq:array_sol_coupled}
    \widetilde{P}_m\left(\tfrac{(2m+1)L}{2},\epsilon\right)&=\widetilde{G}\left(\tfrac{L}{2},\epsilon|0\right)\delta_{m,0}-\widetilde{G}\left(\tfrac{(2m+1)L}{2},\epsilon\Big|\tfrac{(2m+1)L}{2}\right)\nonumber \left[\kappa_-\widetilde{P}_m\left(\tfrac{(2m+1)L}{2},\epsilon\right)-\kappa_+ \widetilde{P}_{m+1}\left(\tfrac{(2m+1)L}{2},\epsilon\right)\right]
    \nonumber \\
    &-\widetilde{G}\left(\tfrac{(2m+1)L}{2},\epsilon\Big|\tfrac{(2m-1)L}{2}\right)\left[\kappa_+ \widetilde{P}_m\left(\tfrac{(2m-1)L}{2},\epsilon\right)-\kappa_- \widetilde{P}_{m-1}\left(\tfrac{(2m-1)L}{2},\epsilon\right)\right],\\[10pt]
    \widetilde{P}_m\left(\tfrac{(2m-1)L}{2},\epsilon\right)&=\widetilde{G}\left(\tfrac{-L}{2},\epsilon|0\right)\delta_{m,0}-\widetilde{G}\left(\tfrac{(2m-1)L}{2},\epsilon\Big|\tfrac{(2m+1)L}{2}\right)\nonumber \left[\kappa_-\widetilde{P}_m\left(\tfrac{(2m+1)L}{2},\epsilon\right)-\kappa_+ \widetilde{P}_{m+1}\left(\tfrac{(2m+1)L}{2},\epsilon\right)\right]
    \nonumber \\
    &-\widetilde{G}\left(\tfrac{(2m-1)L}{2},\epsilon\Big|\tfrac{(2m-1)L}{2}\right)\left[\kappa_+ \widetilde{P}_m\left(\tfrac{(2m-1)L}{2},\epsilon\right)-\kappa_- \widetilde{P}_{m-1}\left(\tfrac{(2m-1)L}{2},\epsilon\right)\right].
\end{align}
After renaming, for ease of notation, $\widetilde{P}_m\left(\tfrac{(2m\pm1)L}{2},\epsilon\right)$ as $\widetilde{Q}_m^r(\epsilon)$ and $\widetilde{Q}_m^l(\epsilon)$, respectively, to indicate with the superscripts whether the position is at the right or left end (i.e. $(2m+1)L/2$ or $(2m-1)L/2$) of the $m$th compartment, see Fig. (\ref{fig:perm_array}) of the main text, we use Eq. (\ref{eq:two_ref_greens}) to obtain 
\begin{align}
    \widetilde{Q}_m^r(\epsilon)=\frac{\text{cosech}\left(\tfrac{L}{2}\sqrt{\tfrac{\epsilon}{S}}\right)}{2\sqrt{D\epsilon}}\delta_{m,0}-\frac{\text{coth}\left(L\sqrt{\tfrac{\epsilon}{D}}\right)}{\sqrt{D \epsilon}}\left[\kappa_-\widetilde{Q}_m^r(\epsilon)-\kappa_+ \widetilde{Q}_{m+1}^l(\epsilon)\right]-\frac{\text{cosech}\left(L\sqrt{\tfrac{\epsilon}{D}}\right)}{\sqrt{D \epsilon}}\left[\kappa_+\widetilde{Q}_m^l(\epsilon)-\kappa_- \widetilde{Q}_{m-1}^r(\epsilon)\right],\label{eq:array_sol_coupled_r}
    \\[5pt]
    \widetilde{Q}_m^l(\epsilon)=\frac{\text{cosech}\left(\tfrac{L}{2}\sqrt{\tfrac{\epsilon}{S}}\right)}{2\sqrt{D\epsilon}}\delta_{m,0}-\frac{\text{cosech}\left(L\sqrt{\tfrac{\epsilon}{D}}\right)}{\sqrt{D \epsilon}}\left[\kappa_-\widetilde{Q}_m^r(\epsilon)-\kappa_+ \widetilde{Q}_{m+1}^l(\epsilon)\right]-\frac{\text{coth}\left(L\sqrt{\tfrac{\epsilon}{D}}\right)}{\sqrt{D \epsilon}}\left[\kappa_+\widetilde{Q}_m^l(\epsilon)-\kappa_- \widetilde{Q}_{m-1}^r(\epsilon)\right].\label{eq:array_sol_coupled_l}
\end{align}

One can see that from Eqs. (\ref{eq:array_sol_coupled_r}) and (\ref{eq:array_sol_coupled_l}) the $m$ dependence is only in the $\widetilde{Q}^r_m(\epsilon)$ and $\widetilde{Q}^l_m(\epsilon)$ functions, ad thus we may utilize the discrete Fourier transform to solve these coupled equations. Taking, $\widetilde{\mathbf{Q}}_m=\left(\widetilde{Q}^r_m(\epsilon),\widetilde{Q}^l_m(\epsilon)\right)^\top$, and defining the discrete Fourier transform as $\widetilde{\mathbf{Q}}(\omega,\epsilon)=\sum_{m=-\infty}^\infty \widetilde{\mathbf{Q}}_m(\epsilon)e^{-i m \omega}$, such that $\sum_{m=-\infty}^\infty \widetilde{\mathbf{Q}}_{m\pm1}(\epsilon)e^{-i m \omega}=e^{\pm i \omega}\widetilde{\mathbf{Q}}(\omega,\epsilon)$, we discrete Fourier transform Eqs. (\ref{eq:array_sol_coupled_r}) and (\ref{eq:array_sol_coupled_l}), and write in matrix form, giving the solution
\begin{align}\label{eq:Q_ex}
    \widetilde{\mathbf{Q}}(\omega,\epsilon)=
    \begin{pmatrix}
        \frac{e^{i \omega } \left[\sqrt{D \epsilon } \text{cosech}\left(\frac{1}{2} L \sqrt{\frac{\epsilon }{D}}\right)+\kappa_+ \left(1+e^{i \omega }\right) \text{sech}\left(\frac{1}{2} L \sqrt{\frac{\epsilon }{D}}\right)\right]}{2 \sqrt{D \epsilon } \text{cosech}\left(L \sqrt{\frac{\epsilon }{D}}\right) \left[e^{i \omega } (\kappa_-+\kappa_+) \cosh \left(L \sqrt{\frac{\epsilon }{D}}\right)-\kappa_--\kappa_+ e^{2 i \omega }\right]+2 D e^{i \omega } \epsilon }\\[0.5cm]
        \frac{e^{i \omega } \sqrt{D \epsilon } \text{cosech}\left(\frac{1}{2} L \sqrt{\frac{\epsilon }{D}}\right)+\kappa_- \left(1+e^{i \omega }\right) \text{sech}\left(\frac{1}{2} L \sqrt{\frac{\epsilon }{D}}\right)}{2 \sqrt{D \epsilon } \text{cosech}\left(L \sqrt{\frac{\epsilon }{D}}\right) \left[e^{i \omega } (\kappa_-+\kappa_+) \cosh \left(L \sqrt{\frac{\epsilon }{D}}\right)-\kappa_--\kappa_+ e^{2 i \omega }\right]+2 D e^{i \omega } \epsilon }
    \end{pmatrix}
\end{align}

To find $\widetilde{P}_m(x,\epsilon)$, all one needs to do is to find the inverse discrete Fourier transform, $\omega\to m$ via 
\begin{equation}\label{eq:inv_fourier}
    \widetilde{\mathbf{Q}}_m(\epsilon)=\frac{1}{2\pi}\int_{-\pi}^\pi \widetilde{\mathbf{Q}}(\omega,\epsilon)e^{i m \omega}d\omega.
\end{equation}
As this integral is difficult to compute, we make the transformation, $e^{i \omega}=z$, which leads to,
\begin{align}\label{eq:inverse_z}
    \widetilde{\mathbf{Q}}_m(\epsilon)=\frac{1}{2\pi} \oint_{|z|\leq 1} z^{m-1}&\left[2 \sqrt{D \epsilon } \text{cosech}\left(L \sqrt{\frac{\epsilon }{D}}\right) \left(z (\kappa_-+\kappa_+) \cosh \left(L \sqrt{\frac{\epsilon }{D}}\right)-\kappa_--\kappa_+ z^2\right)+2 D z \epsilon\right]^{-1}\nonumber\\
    &\times
    \begin{pmatrix}
        z \left[\kappa_+ (z+1) \text{sech}\left(\frac{1}{2} L \sqrt{\frac{\epsilon }{D}}\right)+\sqrt{D \epsilon } \text{cosech}\left(\frac{1}{2} L \sqrt{\frac{\epsilon }{D}}\right)\right]
        \\[0.5cm]
        \kappa_- (z+1) \text{sech}\left(\frac{1}{2} L \sqrt{\frac{\epsilon }{D}}\right)+z \sqrt{D \epsilon } \text{cosech}\left(\frac{1}{2} L \sqrt{\frac{\epsilon }{D}}\right)
    \end{pmatrix}
    dz,
\end{align}
where the contour integral is run counterclockwise.
To make use of Cauchy's residue theorem, we find the poles of the integrand of Eq. (\ref{eq:inverse_z}). As the denominator is a simple quadratic polynomial in $z$, we may write Eq. (\ref{eq:inverse_z}) as follows,
\begin{align}\label{eq:inverse_z_new}
    \widetilde{\mathbf{Q}}_m(\epsilon)=\frac{1}{2\pi} \oint_{|z|\leq 1} \frac{z^{m-1}} {(z-\beta)(z-\gamma)}
    \begin{pmatrix}
        \mathcal{A}_r(z)\\
        \mathcal{A}_l(z)
    \end{pmatrix}
    dz,
\end{align}
where the analytic functions, $\mathcal{A}_r(z)$ and $\mathcal{A}_l(z)$, are given by, 
\begin{align}
    \mathcal{A}_r(z)&=-\frac{z}{\kappa_+\sqrt{D\epsilon}} \left[\kappa_+ (z+1) \sinh\left(\frac{L}{2}  \sqrt{\frac{\epsilon }{D}}\right)+\sqrt{D \epsilon } \cosh\left(\frac{L}{2} \sqrt{\frac{\epsilon }{D}}\right)\right],\label{eq:ar}\\
    \mathcal{A}_l(z)&=-\frac{1}{\kappa_+\sqrt{D\epsilon}} \left[\kappa_- (z+1) \sinh\left(\frac{L}{2}  \sqrt{\frac{\epsilon }{D}}\right)+z\sqrt{D \epsilon } \cosh\left(\frac{L}{2} \sqrt{\frac{\epsilon }{D}}\right)\right],\label{eq:al}
\end{align}
and the roots of the polynomial are, 
\begin{align}
    \beta=&\frac{\sinh \left(L \sqrt{\frac{\epsilon }{D}}\right)}{2 \kappa_+ \sqrt{D \epsilon }} \Bigg\{D\epsilon+\sqrt{D \epsilon } (\kappa_-+\kappa_+) \coth \left(L \sqrt{\frac{\epsilon }{D}}\right)\nonumber\\
    &-\sqrt{D \epsilon  \left((\kappa_-+\kappa_+)^2 \coth ^2\left(L \sqrt{\frac{\epsilon }{D}}\right)+2 \sqrt{D \epsilon } (\kappa_-+\kappa_+) \coth \left(L \sqrt{\frac{\epsilon }{D}}\right)-4 \kappa_- \kappa_+ \text{cosech}^2\left(L \sqrt{\frac{\epsilon }{D}}\right)+D \epsilon \right)} \Bigg\},\\[0.5cm]
    \gamma=&\frac{\sinh \left(L \sqrt{\frac{\epsilon }{D}}\right)}{2 \kappa_+ \sqrt{D \epsilon }} \Bigg\{D\epsilon+\sqrt{D \epsilon } (\kappa_-+\kappa_+) \coth \left(L \sqrt{\frac{\epsilon }{D}}\right)\nonumber\\
    &+\sqrt{D \epsilon  \left((\kappa_-+\kappa_+)^2 \coth ^2\left(L \sqrt{\frac{\epsilon }{D}}\right)+2 \sqrt{D \epsilon } (\kappa_-+\kappa_+) \coth \left(L \sqrt{\frac{\epsilon }{D}}\right)-4 \kappa_- \kappa_+ \text{cosech}^2\left(L \sqrt{\frac{\epsilon }{D}}\right)+D \epsilon \right)} \Bigg\}.
\end{align}

Firstly, we need to identify which poles lie in the region $|z|\leq1$. One can numerically verify that $|\beta|\leq1$ and $|\gamma|>1$ $\forall \kappa_-,\kappa_+,L,D,\Re(\epsilon)>0$. The condition on $\epsilon$ is due to the only singularities from $\epsilon$ being at $\epsilon=0$, thus when performing the inverse Laplace transform, $\epsilon\to t$, one has the condition $\Re(\epsilon)>0$. 

So, we have the simple pole at $z=\beta$, which for $\widetilde{Q}_m^r(\epsilon)$ is the only pole for $m>-1$, whilst for $\widetilde{Q}_m^l(\epsilon)$ is the only pole for $m>0$, this is due to the different forms of the elements in the column vector in Eq. (\ref{eq:inverse_z}) with the $z$ terms contributing to the singularity of $z^{m-1}$. For the pole at $z=\beta$, the residues are simply,
\begin{equation}\label{eq:res_bet}
   \text{Res}_{z=\beta}=\frac{\beta^{m-1}}{\beta-\gamma}
    \begin{pmatrix}
        \mathcal{A}_r(\beta) \\
        \mathcal{A}_l(\beta). 
    \end{pmatrix}
\end{equation}

The other pole is located at $z=0$ for certain values of $z$, to find the residue we look for the Laurent series. By writing, 
\begin{equation}
    \frac{1}{(z-\beta)(z-\gamma)}=\frac{1}{\gamma(\beta-\gamma)}\frac{1}{1-z/\gamma}-\frac{1}{\beta(\beta-\gamma)}\frac{1}{1-z/\beta}=\frac{1}{\beta-\gamma}\sum_{k=0}^\infty z^k \left(\gamma^{-1-k}- \beta^{-1-k}\right), 
\end{equation}
after combining with the integrand in Eq. (\ref{eq:inverse_z_new}), we find the residue by looking for the factor multiplying the $z^{-1}$ term in the Laurent series, which after considering the $z$ dependence in $\mathcal{A}_r(z)$ and $\mathcal{A}_l(z)$, we find the residues to be 
\begin{equation}\label{eq:res_zero}
    \text{Res}_{z=0}=\left\{\frac{\gamma^{m-1}}{\beta-\gamma} 
    \begin{pmatrix}
        \mathcal{A}_r(\gamma)\\
        \mathcal{A}_l(\gamma)
    \end{pmatrix}
    -\frac{\beta^{m-1}}{\beta-\gamma}
    \begin{pmatrix}
        \mathcal{A}_r(\beta)\\
        \mathcal{A}_l(\beta)
    \end{pmatrix}
    \right\} 
    \begin{pmatrix}
        \Theta(-m-1)\\
        \Theta(-m)
    \end{pmatrix}
\end{equation}
where the column vector of $\Theta(z)$ is due to the requirement of $m$ in the $z^m-1$ term and the different $z$ dependence of $\mathcal{A}_r(z)$ and $\mathcal{A}_l(z)$ in Eqs. (\ref{eq:ar}) and (\ref{eq:al}), and the Heaviside step function, $\Theta(z)$, is defined as,
\begin{equation}
    \Theta(z)=\left\{
    \begin{array}{ll}
         1, \ z\geq0,\\
         0, \ z<0.
    \end{array}
    \right.
\end{equation}
After summing the residues in Eqs. (\ref{eq:res_bet}) and (\ref{eq:res_zero}) together, we find $\widetilde{\mathbf{Q}}_m(\epsilon)$ as
\begin{equation}\label{eq:Q_final}
    \widetilde{\mathbf{Q}}_m(\epsilon)=\frac{\beta^{m-1}}{\beta-\gamma}
    \begin{pmatrix}
        \mathcal{A}_r(\beta) \Theta(m)\\
        \mathcal{A}_l(\beta) \Theta(m-1)
    \end{pmatrix}
    +\frac{\gamma^{m-1}}{\beta-\gamma}
    \begin{pmatrix}
        \mathcal{A}_r(\gamma) \Theta(-m-1)\\
        \mathcal{A}_l(\gamma) \Theta(-m)
    \end{pmatrix}
    .
\end{equation}
Equation (\ref{eq:Q_final}) has been verified by comparing it to the numerically solved integral in Eq. (\ref{eq:inv_fourier}).

Finally, recalling that $\widetilde{P}_m\left(\tfrac{(2m+1)L}{2},\epsilon\right)=\widetilde{Q}_m^r(\epsilon)$ and $\widetilde{P}_m\left(\tfrac{(2m-1)L}{2},\epsilon\right)=\widetilde{Q}_m^l(\epsilon)$, and from Eq. (\ref{eq:array_sol}) in the main text, we may write the form of $\widetilde{P}_m(x,\epsilon)$ as follows,
\begin{align}\label{eq:array_exact_sol_vec}
    \widetilde{P}_m(x,\epsilon)&=\widetilde{G}(x,\epsilon|0)\delta_{m,0} - \left(\kappa_-\widetilde{G}\left(x,\epsilon|(2m+1)L/2\right), \kappa_+ \widetilde{G}\left(x,\epsilon|(2m-1)L/2\right) \right) \widetilde{\mathbf{Q}}_m(\epsilon) \nonumber \\
    &+ \left(0,\kappa_+\widetilde{G}\left(x,\epsilon|(2m+1)L/2\right)\right) \widetilde{\mathbf{Q}}_{m+1}(\epsilon)+ \left(\kappa_-\widetilde{G}\left(x,\epsilon|(2m-1)L/2\right),0\right) \widetilde{\mathbf{Q}}_{m-1}(\epsilon),
\end{align}
where the Green's functions are defined in Eqs. (\ref{eq:two_ref_greens}). Clearly the inverse Laplace transform is not feasible analytically, but we can investigate the moments in the long-time limit. When $\kappa_+=\kappa_-$  a proposed time dependent form is presented in Ref. \cite{powles1992exact}, however due to the lack of a derivation, and the presence in some of the infinite summation of terms containing the word `step 2' that lack any meaning in continuous space-time setting, we have no means to verify that it represents the inverse Laplace transform of Eq. (\ref{eq:array_exact_sol_vec}) for symmetric barriers. We are thus drawn to concur with the authors of ref. \cite{slkezak2021diffusion} casting doubts on the validity of the proposed $P_m(x,t)$ in ref. \cite{powles1992exact}.

\subsection{First moment across all compartments}

Here we use Eq. (\ref{eq:array_exact_sol_vec}) to find the first moment, the mean $M_1(t)=\langle X(t)\rangle$, of a Brownian particle in the presence of an infinite array of periodically placed identical asymmetric barriers. For this set-up the Laplace transform of the mean can be found using $\widetilde{P}_m(x,\epsilon)$, via
\begin{equation}\label{eq:mean}
    \widetilde{M}_1(\epsilon)=\sum_{m=-\infty}^\infty \int_{\frac{2m-1}{2}L}^{\frac{2m+1}{2}L} x \widetilde{P}_m(x,\epsilon) dx.
\end{equation}
Considering Eq. (\ref{eq:array_exact_sol_vec}) the integral in Eq. (\ref{eq:mean}) is only over the Green's functions in Eq. (\ref{eq:two_ref_greens}), and this integral can be computed easily. As we are interested in extracting effective transport parameters from the long-time form of the moments, we expand around $\epsilon\to0$ to find
\begin{equation}
    \int_{\frac{2m-1}{2}L}^{\frac{2m+1}{2}L} x \widetilde{G}\left(x,\epsilon|(2m\pm1)L/2\right)\mathrel{\underset{\epsilon \to 0}{\sim}} \frac{L m}{\epsilon}.
\end{equation}
Using this with Eq. (\ref{eq:array_exact_sol_vec}), with the first term on the right-hand side vanishing, and performing the summation leads to, 
\begin{equation}\label{eq:mean_sum}
    \widetilde{M}_1(\epsilon)\mathrel{\underset{\epsilon \to 0}{\sim}} \frac{L \left(\kappa _- \left((\beta -1) \beta  \mathcal{A}_r(\gamma)-(\gamma -1) \gamma  \mathcal{A}_r(\beta)\right)+\beta  \gamma  \kappa _+ \left((\gamma -1) \mathcal{A}_l(\beta)-(\beta -1) \mathcal{A}_l(\gamma)\right)\right)}{(\beta -1) \beta  (\gamma -1) \gamma  \epsilon },
\end{equation}
where the dependence of $\epsilon$ on the parameters, $\mathcal{A}_r, \mathcal{A}_l, \beta, \gamma$, is suppressed. Now, as we are interested in the $t\to \infty$ case, we insert $\mathcal{A}_r, \mathcal{A}_l, \beta, \gamma$ into Eq. (\ref{eq:mean_sum}), expand around $\epsilon\to 0$ to first order to give,
\begin{equation}
    \widetilde{M}_1(\epsilon)\mathrel{\underset{\epsilon \to 0}{\sim}} \frac{2 D (\kappa_--\kappa_+)}{\epsilon ^2 (2 D+L (\kappa_-+\kappa_+))},
\end{equation}
which after Laplace transforming gives,
\begin{equation}
    M_1(t)\mathrel{\underset{t \to \infty}{\sim}} \nu_\text{eff} t,
\end{equation}
for $\nu_\text{eff}$ as in Eq. (\ref{eq:eff_vel}) in the main text.

\subsection{Second moment across all compartments}

To find the second moment, $M_2(t)=\langle X^2(t) \rangle$, we follow a very similar procedure as finding, $M_1(t)$, due to the second moment being defined as, 
\begin{equation}
    \widetilde{M}_2(\epsilon)=\sum_{m=-\infty}^\infty \int_{\frac{2m-1}{2}L}^{\frac{2m+1}{2}L} x^2 \widetilde{P}_m(x,\epsilon) dx.
\end{equation}
As before this integral is only over the Green's functions and in the long-time limit, we have,
\begin{equation}
    \int_{\frac{2m-1}{2}L}^{\frac{2m+1}{2}L} x^2 \widetilde{G}\left(x,\epsilon|(2m\pm1)L/2\right)\mathrel{\underset{\epsilon \to 0}{\sim}} \frac{L^2 m^2}{\epsilon }+\frac{L^2}{12 \epsilon }.
\end{equation}
Then using Eq. (\ref{eq:array_exact_sol_vec}) we may perform the summation to give, 
\begin{align}\label{eq:var_sum}
    \widetilde{M}_2(\epsilon)&\mathrel{\underset{\epsilon \to 0}{\sim}} L^2\Big\{ \kappa _- \Big[(\beta +1) (\gamma -1)^2 \gamma  \mathcal{A}_r(\beta)-(\beta -1)^2 \beta  (\gamma +1) \mathcal{A}_r(\gamma)\Big]\\
    &+\beta  \gamma  \kappa _+ \Big[(\beta -1)^2 (\gamma +1) \mathcal{A}_l(\gamma)-(\beta +1) (\gamma -1)^2 \mathcal{A}_l(\beta)\Big]\Big\}\Big[(\beta -1)^2 \beta  (\gamma -1)^2 \gamma  \epsilon \Big]^{-1}.
\end{align}
Inserting $\mathcal{A}_r, \mathcal{A}_l, \beta, \gamma$ into Eq. (\ref{eq:var_sum}) and expanding around $\epsilon\to 0$ to second order, we arrive at,
\begin{align}
    \widetilde{M}_2(\epsilon)&\mathrel{\underset{\epsilon \to 0}{\sim}} \frac{8 D^2 (\kappa_--\kappa_+)^2}{\epsilon ^3 \Big[2 D+L (\kappa_-+\kappa_+)\Big]^2} \\
    &+ \frac{D L \Big[24 D^2 (\kappa_-+\kappa_+)+2 D L \left(7 \kappa_-^2+34 \kappa_- \kappa_++7 \kappa_+^2\right)+L^2 (\kappa_-+\kappa_+) \left(5 \kappa_-^2+14 \kappa_- \kappa_++5 \kappa_+^2\right)\Big]}{3 \epsilon ^2 \Big[2 D+L (\kappa_-+\kappa_+)\Big]^3}.
\end{align}
After Laplace transforming, we have
\begin{equation}
    M_2(t)\mathrel{\underset{t \to \infty}{\sim}} 2  D_\text{eff} t + \nu_\text{eff}^2 t^2,
\end{equation}
where $D_\text{eff}$ is defined in Eq. (\ref{eq:diff_eff}) of the main text.

\end{widetext}

\bibliography{references}
\end{document}